\documentclass[runningheads]{llncs}
\usepackage[utf8]{inputenc}
\usepackage[T1]{fontenc}
\usepackage[all]{xy}
\usepackage{wrapfig}
\usepackage{graphicx}
\usepackage{booktabs}
\usepackage{fancyvrb}
\usepackage{relsize}
\usepackage[bookmarks=true,bookmarksnumbered=true]{hyperref}
\usepackage[all]{hypcap}
\usepackage{listings}

\usepackage[english]{babel}
\usepackage{csquotes}

\newenvironment{smallverbatim}
{\Verbatim[fontsize=\footnotesize, frame=single, samepage=true, tabsize=8, commandchars=\\\{\}]}
{\endVerbatim}

\newif \ifllncs
\llncstrue

\def \mytitle {High-Level Multi-Threading in hProlog}
\def \myauthor {Timon Van Overveldt}
\def \demoen {Bart Demoen}
\def \mydept {Department of Computer Science}
\def \myuniv {Katholieke Universiteit Leuven, Belgium}
\def \myemail {timon.vanoverveldt@student.kuleuven.be}
\def \demoenemail {bart.demoen@cs.kuleuven.be}

\hypersetup{
	pdftitle=\mytitle,
	pdfauthor=\myauthor
}

\ifllncs
\title{\mytitle}
\else
\title{\textbf{\mytitle}}
\fi

\ifllncs
\author{
	\myauthor\thanks{This work is part of the first author's bachelor's thesis.}
	\and
	\demoen
}
\institute{
	\mydept\\
	\myuniv\\
	\href{mailto:\myemail}{\myemail}\\
	\href{mailto:\demoenemail}{\demoenemail}
}
\else
\author{
	\myauthor\\
	\mydept\\
	\myuniv\\
	\href{mailto:\myemail}{\myemail}
}
\fi

\date{}

\begin{document}

\maketitle

\begin{abstract}
A new high-level interface to multi-threading in Prolog, implemented in hProlog, is described.
Modern CPUs often contain multiple cores and through high-level multi-threading a programmer can leverage this power without having to worry about low-level details.
Two common types of high-level explicit parallelism are discussed: independent and-parallelism and competitive or-parallelism.
A new type of explicit parallelism, pipeline parallelism, is proposed.
This new type can be used in certain cases where independent and-parallelism and competitive or-parallelism cannot be used.
\end{abstract}

\section{Introduction}
Modern CPUs often have multiple cores and thus are capable of executing multiple threads concurrently.
This makes fully exploiting the processing power of CPUs a non-trivial problem.
There are various Prolog implementations with support for multi-threading.
Some systems have implemented implicit parallelism \cite{springerlink:10.1007/3-540-48159-1_13,springerlink:10.1007/BF01407834,matsphd}, where general Prolog programs are automatically parallelised by either concurrently executing multiple goals or concurrently exploring multiple branches in the problem tree.
However, these implementations turned out to be hard to maintain and they also make it hard to control the degree of parallelism.

Another way to exploit multi-threading is through explicit parallelism.
With explicit parallelism, the programmer specifies exactly how the program should be parallelised.
This specification can vary in granularity.
For example, some implementations use a low-level interface based on POSIX threads \cite{DBLP:conf/iclp/Wielemaker03,xsb_prolog_manual,yap_manual}.
These implementations allow a great deal of control over the underlying threads.
However, since these interfaces originate from a procedural API originally written for the C programming language, they are not very declarative.
Other implementations have gone a different route, providing high-level interfaces tailored to specific types of problems.
It is hard to classify implementations strictly based on this distinction, since a number of implementations provide both low-level and high-level interfaces.

Two types of parallelism are \emph{competitive or-parallelism} \cite{springerlink:10.1007/978-3-540-89982-2_63} and \emph{independent and-parallelism} \cite{DBLP:journals/ngc/HermenegildoG91,Moura:2008:HMP:1785754.1785772}.
With the former, a disjunction of alternative goals is executed concurrently by letting each goal compete to provide the solution to a single problem.
The first solution to become available is used and the remaining goals are stopped.
With the latter, each goal in a conjunction of independent goals is executed concurrently.
This type of parallelism is not suited for conjunctions where goals depend on each other.

In this paper, we describe a moderately high-level interface with which one can implement high-level problem-specific predicates.
Our interface is inspired by the interface found in LeanProlog \cite{Tarau:2011:CPC:1926354.1926364} and tries to relieve the programmer of having to care for low-level details, allowing him to focus on the problems at hand.
On the other hand, we have tried to make our interface general enough to be suitable for a variety of problem types.
The interface can be situated in between high-level interfaces specific to certain types of problems, and the low-level POSIX interface.
We have implemented the API in hProlog, a Prolog implementation written in C and the successor to dProlog \cite{wamvariations}.
The implementation required relatively few architectural changes, mainly synchronisation of some data areas and making other data areas thread-local.

We describe our interface and its related concepts in \autoref{sec:threads}.
In \autoref{sec:competitive_or} and \autoref{sec:independent_and} we show how our interface can be used to implement competitive or-parallelism and independent and-parallelism.
\autoref{sec:pipeline} describes a new type of parallelism which we call \emph{pipeline parallelism}.
In \autoref{sec:performance} results from benchmarks testing the performance of our implementation with the three discussed types of parallelism are presented.
Finally, in \autoref{sec:related_work} we compare our interface and implementation with other implementations and in \autoref{sec:conclusion} we formulate our conclusion.

\section{Multi-Threading Support}
\label{sec:threads}
The language constructs we describe are centered around \emph{threads}, identified by opaque terms called \emph{thread IDs}.
Each thread concurrently executes a goal in a separate Prolog engine.
A thread is automatically terminated after generating all solutions to its goal.
The solutions are made available as soon as they are generated.
\subsection{Spawning Threads}
Threads are spawned using the \texttt{spawn/3} predicate.
\begin{smallverbatim}
spawn(AnswerPattern, Goal, ID)
\end{smallverbatim}
\texttt{spawn/3} spawns a new thread executing a copy of \texttt{Goal}.
On success of the call to \texttt{spawn/3}, \texttt{ID} is unified with the thread's ID.
\texttt{AnswerPattern} is a term that can contain variables from \texttt{Goal}.

Threads share as much memory areas as possible.
The only areas that are thread-private are the trail, the local stack, the global stack, and the choice point stack.
All these data areas are expanded as needed.
At the lower level, Prolog threads are mapped to POSIX threads.
Thus, all Prolog threads are scheduled by the operating system.

\subsection{Message Passing}
\begin{sloppypar}
Each time a thread generates a solution to its goal, a \emph{copy} of the term \texttt{the(AnswerPattern)} is sent to the thread's \emph{default recipient}, which in general is the thread calling \texttt{spawn/3}.
Using an answer pattern, one can specify exactly which variables need to be returned to the default recipient, minimising the overhead of copying answers.
\end{sloppypar}

After a thread has generated all solutions to its goal, the \texttt{no} atom is sent to the default recipient to notify it of the threads termination.
Because answers are wrapped in \texttt{the/1} and termination messages are not, a recipient can easily differentiate between the two.

Messages can also be sent explicitly:
\begin{smallverbatim}
send(Term)
\end{smallverbatim}
sends a copy of \texttt{the(Term)} to the default recipient, while
\begin{smallverbatim}
send(ID, Term)
\end{smallverbatim}
sends the copy to the thread identified by \texttt{ID}.
Threads receive messages by calling \texttt{receive}:
\begin{smallverbatim}
receive(Term)
receive(ID, Term)
\end{smallverbatim}
\texttt{receive/1} consumes the first message in the thread's inbox, regardless of message's sender.
When \texttt{ID} is free, \texttt{receive/2} unifies it with the ID of the message's sender.
If there are no messages, the call to \texttt{receive/1} or \texttt{receive/2} blocks until a message becomes available.
If \texttt{ID} is a valid thread ID, the first message sent by that thread is consumed.
Currently, both \texttt{send} and \texttt{receive} fail if the thread ID is invalid.
Because every message is a copy of a term, further alterations to an already sent term do not propagate to its copy.
This eliminates the need for synchronisation and simplifies the design of the message passing API.

\subsection{Terminating Threads}
\label{subsec:terminating_threads}
A thread can be (preemptively) terminated by issuing a call to
\begin{smallverbatim}
stop(ID)
\end{smallverbatim}
Every thread may terminate another thread.
When a thread is terminated, the threads it has spawned are \emph{not} terminated as well.

The virtual machine machine contains several checkpoints where the cancellation of a thread is checked.
These checkpoints are located in the same places as where heap overflows are checked.
Because cancellation is not immediate, a thread may continue executing for a short time until such a checkpoint is reached.
After reaching a cancellation checkpoint, resources are freed and the thread is shut down.
Terminating a thread is an \emph{asynchronous} operation: the call to \texttt{stop/1} returns without waiting for the thread to fully shut down.
After stopping a thread, its ID must not be used anymore.
As a temporary measure, any messages from the terminated thread remaining in the calling thread's inbox are purged.
What happens to other messages remaining in other threads' inboxes is currently undefined.

\subsection{Hubs}
\emph{Hubs} are message queues \cite{DBLP:journals/corr/abs-1102-1178}, identified by a \emph{hub ID}.
They are created by calling \texttt{hub/1}.
\begin{smallverbatim}
hub(HubID)
\end{smallverbatim}
By specifying the hub's ID, messages can be sent to and received from it.
\begin{smallverbatim}
send(HubID, Term)
receive(HubID, Term)
receive(HubID, ThreadID, Term)
\end{smallverbatim}
In this case \texttt{receive/2} receives a message from the hub specified by \texttt{HubID}, regardless of the message's sender, while \texttt{receive/3} receives a message from the hub specified by \texttt{HubID}.
If \texttt{ThreadID} is a valid thread ID, the first message sent by that thread is consumed.
Otherwise, if \texttt{ThreadID} is unbound, it is unified with the ID of the message's sender.

Threads can be linked to a hub: when the hub is terminated, all of its linked threads are automatically and \emph{synchronously} terminated.
When a thread is linked to a hub, the hub becomes the thread's default recipient.
As such, the solutions the thread generates, as well as its termination message, are sent to the hub it is linked to.
To spawn a thread linked to a hub
\begin{smallverbatim}
spawn_link(HubID, AnswerPattern, Goal, ID)
\end{smallverbatim}
is used, where \texttt{HubID} is the ID of the hub.

As mentioned, when a hub is stopped using \texttt{stop/1}, all linked threads are also stopped.
The call to \texttt{stop/1} is synchronous: it blocks until all the linked threads have fully shut down.
As such, after stopping a hub, one can be sure that the resources of any linked threads have been freed.
%
%We have also implemented support for linking threads directly.
%However, we will not discuss this type of linking any further.

\subsection{Limitations}
\label{subsec:limitations}
Currently, there is a limit on the number of threads that can be simultaneously executed.
If desired, this limit can be specified at run time.
There are as many thread IDs as the maximum number of simultaneous threads and these IDs are recycled after they are discarded.
However a thread's ID cannot be discarded as long as \texttt{stop/1} isn't called, signalling that a thread will not be used anymore.
As such, when a large number of threads are created but never stopped, one can run out of thread IDs, even if these threads are not executed simultaneously.

Currently, there is no explicit support for synchronisation between threads.
However, we expect that implementing synchronisation constructs will not require drastic changes.
\\

\noindent
The next sections discuss three types of parallellism and shows how they can be implemented using the language constructs described earlier.

\section{Competitive Or-Parallelism}
\label{sec:competitive_or}
The concept of \emph{competitive or-parallelism}, as described in \cite{springerlink:10.1007/978-3-540-89982-2_63},
``is based on the interpretation of an explicit \emph{disjunction} of subgoals as a set of concurrent alternatives, each running in its own thread''.
The subgoals, each implementing a different algorithm, compete to provide the solution to the problem.
When a solution is found, the remaining subgoals are stopped.

Competitive or-parallelism is useful when there exist multiple alternative algorithms to solve a single problem.
A given algorithm might perform better for one instance of the problem, while it might perform worse for another.
As such, the performance of an algorithm depends on the specifics of the problem.
Alternatively, an algorithm may never terminate, requiring the use of competitive or-parallelism to guarantee that a solution is always found.

Logtalk \cite{Moura:2008:HMP:1785754.1785772} provides a \texttt{threaded/1} predicate for competitive or-parallelism.
It accepts a disjunction of goals and runs these concurrently.
The first solution is returned to the user by binding the variables in the disjunction.
The remaining threads are automatically terminated.

The following example, adapted from \cite{springerlink:10.1007/978-3-540-89982-2_63}, solves the 
\emph{water jugs problem}\footnote{Given several jugs of different capacities, we want to measure a certain amount of water. Jugs can be filled or emptied or their contents can be transferred to another jug.}
using three competing algorithms: breadth first, depth first and hill climbing.
\ifllncs
\begin{smallverbatim}
solve(Jugs, Moves) :-
	threaded((
		  breadth_first_solve(Jugs, Moves)
		; depth_first_solve(Jugs, Moves)
		; hill_climbing_solve(Jugs, Moves)
	)).
\end{smallverbatim}
\else
\begin{smallverbatim}
solve(Jugs, Moves) :-
	threaded((
		  bf_solve(Jugs, Moves)
		; df_solve(Jugs, Moves)
		; hc_solve(Jugs, Moves)
	)).
\end{smallverbatim}
\fi
The call to \texttt{threaded/1} is semideterministic and opaque to cuts; there is no backtracking over completed calls.
One can achieve the same result using the language constructs described in \autoref{sec:threads}:
\ifllncs
\begin{smallverbatim}
solve(Jugs, Moves) :-
	hub(H),
	spawn_link(H, Moves, breadth_first_solve(Jugs, Moves), T1),
	spawn_link(H, Moves, depth_first_solve(Jugs, Moves), T2),
	spawn_link(H, Moves, hill_climbing_solve(Jugs, Moves), T3),
	receive(H, the(Moves)),
	stop(H).
\end{smallverbatim}
\else
\begin{smallverbatim}
solve(Jugs, Moves) :-
	hub(H),
	spawn_link(H, Moves, bf_solve(Jugs,Moves), T1),
	spawn_link(H, Moves, df_solve(Jugs,Moves), T2),
	spawn_link(H, Moves, hc_solve(Jugs,Moves), T3),
	receive(H, the(Moves)),
	stop(H).
\end{smallverbatim}
\fi
Note that we have also implemented \texttt{threaded/1} using our language constructs.
However for the sake of brevity, we only show how the same result can be achieved, without showing the full implementation of \texttt{threaded/1}.

\section{Independent And-Parallelism}
\label{sec:independent_and}
The concept of \emph{independent and-parallelism} consists of an explicit \emph{conjunction} of subgoals, each running concurrently.
In contrast to competitive or-parallelism, the subgoals do not compete to provide the solution to a problem.
Instead, the conjunction is interpreted as a set of parallelisable goals, all of which need to succeed for the whole conjunction to succeed.
In order for the goals to be parallelisable, goals cannot in general use each other's output as their input.

The \texttt{threaded/1} predicate described above, also lends itself to independent and-parallelism.
For example, if we want to calculate a given Fibonacci number using two threads, we could use
\begin{smallverbatim}
fibonacci(N, F) :-
	N1 is N-1,
	N2 is N-2,
	threaded((
		do_fibonacci(N1, F1),
		do_fibonacci(N2, F2)
	)),
	F is F1 + F2.
\end{smallverbatim}
As with competitive or-parallelism, the call to \texttt{threaded/1} is semideterministic.
The same result can be achieved using our language constructs described in \autoref{sec:threads}:
\begin{smallverbatim}
fibonacci(N, F) :-
	N1 is N-1,
	N2 is N-2,
	spawn(F1, do_fibonacci(N1, F1), T1),
	spawn(F2, do_fibonacci(N2, F2), T2),
	receive(T1, the(F1)),
	receive(T2, the(F2)),
	stop(T1),
	stop(T2),
	F is F1 + F2.
\end{smallverbatim}
As illustrated, our interface provides everything to implement both competitive or- and independent and-parallelism.

\section{Pipeline Parallelism}
\label{sec:pipeline}
In the previous sections we have described ways to use high-level multi-threading constructs for different types of parallelism.
However, until now these types were restricted to either competitive or-parallelism or independent and-parallelism.
In this section, we describe a new type of parallelism, one which can be used to concurrently execute conjunctions in which goals depend on each other.
When parallelising such conjunctions, each goal needs to be solved in the order in which it appears.
Additionally we want the solutions to the conjunction to appear in the same order as they would when the set of goals was not executed concurrently.
Independent and-parallelism is not suited for use with these types of conjunctions.

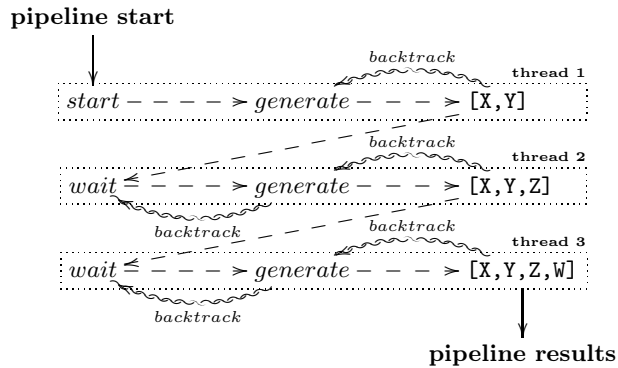
\begin{figure}[t]
\centerline{
{\small
\xymatrix @R=1.5pc
{
% Nodes in matrix
\textbf{pipeline\ start} \\
start & generate & \texttt{[X,Y]}\ \ \ \ \ \  {}\save[]+<0.35cm,0.4cm>*\txt{\tiny{\textbf{thread 1}}}\restore \\
wait & generate & \texttt{[X,Y,Z]}\ \ \  {}\save[]+<0.35cm,0.4cm>*\txt{\tiny{\textbf{thread 2}}}\restore \\
wait & generate & \texttt{[X,Y,Z,W]} {}\save[]+<0.35cm,0.4cm>*\txt{\tiny{\textbf{thread 3}}}\restore \\
 & & \textbf{pipeline\ results}
% Groups
\save "2,1"."2,3"*[F.]\frm{}
\restore
\save "3,1"."3,3"*[F.]\frm{}
\restore
\save "4,1"."4,3"*[F.]\frm{}
\restore
\ar@{->}"1,1";"2,1" % Arrow from start to thread 1
\ar@{-->}"2,1";"2,2" % Thread 1 internal arrow from "start" to "generate"
\ar@{-->}"2,2";"2,3" % Thread 1 internal arrow from "generate" to "partial solution"
\ar@{~>}_{backtrack} @/_1pc/ "2,3";"2,2" % Arrow from "partial solution" to "generate"
\ar@{-->}"2,3";"3,1" % Arrow from thread 1 "generate" to thread 2 "wait"
\ar@{-->}"3,1";"3,2" % Thread 2 internal arrow from "wait" to "generate"
\ar@{-->}"3,2";"3,3" % Thread 2 internal arrow from "generate" to "partial solution"
\ar@{~>}_{backtrack} @/_1pc/ "3,3";"3,2" % Arrow from "partial solution" to "generate"
\ar@{~>}^{backtrack} @/^1pc/ "3,2";"3,1" % Arrow from "generate" to "wait"
\ar@{-->}"3,3";"4,1" % Arrow from thread 1 "generate" to thread 2 "wait"
\ar@{-->}"4,1";"4,2" % Thread 3 internal arrow from "wait" to "generate"
\ar@{-->}"4,2";"4,3" % Thread 3 internal arrow from "generate" to "complete solution"
\ar@{~>}_{backtrack} @/_1pc/ "4,3";"4,2" % Arrow from "partial solution" to "generate"
\ar@{~>}^{backtrack} @/^1pc/ "4,2";"4,1" % Arrow from "generate" to "wait"
\ar@{->}"4,3";"5,3" % Arrow from thread 3 "complete solution" to end
}
}
}
\caption{Pipelined execution of a conjunction of three goals.}
\label{fig:pipeline}
\end{figure}

We propose an approach similar to the instruction pipeline found in modern CPUs.
As an example, we describe the concept using a conjunction of three goals.
\autoref{fig:pipeline} is an illustration for this example.

Parallelising the example 3-goal conjunction \texttt{g1(X,Y), g2(Y,Z), g3(Z,W)} using a pipeline consists of the following stages:
\begin{description}
\item[Preparation]\hfill\newline
A hub is spawned to hold the results of the pipeline.
Three threads linked to the hub are spawned, one for each goal in the conjunction.
\item[First stage]\hfill\newline
The first thread starts executing \texttt{g1(X,Y)}.
It generates a solution with accompanying variable bindings \texttt{[X,Y]} and sends those to the second thread.
After forwarding these bindings, it immediately backtracks to find the next solution to its goal, until all solutions have been found.
\item[Second stage]\hfill\newline
Meanwhile, the second thread waits for bindings \texttt{[X,Y]} from the first thread.
As soon as it receives these bindings it executes \texttt{g2(Y,Z)} using them, forwarding the resulting bindings \texttt{[X,Y,Z]} each time a solution is found. It does this until all bindings from the first thread have been received.
\item[Third stage]\hfill\newline
The third thread behaves just like the second, generating all solutions to \texttt{g3(Z,W)} for each set of bindings \texttt{[X,Y,Z]} it receives from the second thread.
The variable bindings \texttt{[X,Y,Z,W]} generated by this thread are sent to the hub.
\end{description}

\subsection{Pipeline Implementation}
Here we offer a general implementation of a pipeline using the language constructs discussed in \autoref{sec:threads}, as the predicate \texttt{piped/2}.
This predicate can be used as follows:
\ifllncs
\begin{smallverbatim}
?- piped((member(X, [1,2]), member(X, [2,3])), ID),
   stop(ID).
X = 2
\end{smallverbatim}
\else
\begin{smallverbatim}
?- piped((member(X, [1,2]),
          member(X, [2,3])), ID),
   stop(ID).
X = 2
\end{smallverbatim}
\fi
As shown, the first and only solution to the given conjunction is generated, after which the pipeline is stopped.

The predicate \texttt{piped/2} can be implemented as follows:
\begin{smallverbatim}
piped(Goals, ID) :-
	term_variables(Goals, Vars),
	pipe_create(Goals, Vars, Id),
	pipe_results(ID, Vars).
\end{smallverbatim}
First, \texttt{term\_variables/2} is called to extract the variables used in the conjunction.
Since this list of variables is sufficient to fully describe a partial solution, its use reduces overhead by eliminating the forwarding of unnecessary terms between stages.
Next, \texttt{piped/2} creates a new pipe using \texttt{pipe\_create/3}.
\ifllncs
\begin{smallverbatim}
pipe_create(Goals, Vars, End) :-
	hub(End),
	spawn_pipe_stages(End, Vars, Goals, [Head|Stages]),
	link_pipe_stages([Head|Stages], End),
	send(Head, _),
	send(Head, done).
\end{smallverbatim}
\else
\begin{smallverbatim}
pipe_create(Gs, Vars, End) :-
	hub(End),
	spawn_pipe_stages(End, Vars, Gs, [Head|Rest]),
	link_pipe_stages([Head|Rest], End),
	send(Head, _),
	send(Head, done).
\end{smallverbatim}
\fi
\texttt{pipe\_create/3} creates a new pipe from a conjunction of goals, returning the pipe's ID.
\texttt{spawn\_pipe\_stages/4} spawns a thread for every goal in the conjunction.
The threads are linked to the hub, which also acts as the end of the pipe.
The ID of the hub is also used as the ID of the pipeline.
\ifllncs
\begin{smallverbatim}
spawn_pipe_stages(End, Vars, (Goal, Goals), [ID | IDs]) :-
	!, spawn_link(End, [], pipe_stage(Vars, Goal), ID),
	spawn_pipe_stages(End, Vars, Goals, IDs).
spawn_pipe_stages(End, Vars, (Goal), [ID]) :-
	spawn_link(End, [], pipe_stage(Vars, Goal), ID).
\end{smallverbatim}
\else
\begin{smallverbatim}
spawn_pipe_stages(End, Vars, (G,Gs), [ID|IDs]):-
	!,
	spawn_link(End, [], pipe_stage(Vars, G), ID),
	spawn_pipe_stages(End, Vars, Gs, IDs).
spawn_pipe_stages(End, Vars, (G), [ID]) :-
	spawn_link(End, [], pipe_stage(Vars, G), ID).
\end{smallverbatim}
\fi
In turn, \texttt{link\_pipe\_stages/2} links together the stages in the pipe by sending each thread the ID of the stage following it.
\begin{smallverbatim}
link_pipe_stages([Stage], End) :-
	!, send(Stage, End).
link_pipe_stages([Stage, Next | IDs], End) :-
	send(Stage, Next),
	link_pipe_stages([Next | IDs], End).
\end{smallverbatim}
Each thread is started with \texttt{pipe\_stage/2} as its start goal.
This predicate handles the forwarding of (partial) solutions, backtracking and the termination of each stage in the pipeline.
\begin{smallverbatim}
pipe_stage(Vars, Goal) :-
	receive(the(Next)),
	repeat,
	receive(_, the(In)),
	(
	  In == done -> send(Next, done), !, fail
	; Vars = In
	),
	Goal,
	send(Next, Vars),
	fail.
\end{smallverbatim}
This is everything we need to create a pipeline.
Note that in \texttt{pipe\_create/3}, after spawning and linking the stages of the pipeline, a dummy variable and the message \texttt{done} are immediately sent to the head of the pipe to make it start generating answers.

Finally, after creating and starting a new pipe, \texttt{piped/2} calls \texttt{pipe\_results/2}.
This is a backtrackable predicate that unifies \texttt{Vars} with a result from the pipeline until no more results are available.
\begin{smallverbatim}
pipe_results(End, Vars) :-
	repeat,
	receive(End, In),
	(
	  In == the(done) -> !, fail
	; the(Vars) = In
	).
\end{smallverbatim}

Note that \texttt{piped/2} returns an ID.
As previously mentioned, this ID is actually the ID of the hub to which all stages in the pipeline are linked.
As such, calling \texttt{stop/1} on the ID terminates the whole pipeline.

\subsection{A Pipelined Findall}
As an example, we show a version of \texttt{findall/3} that executes its goals using pipeline parallelism, while retaining the semantics of \texttt{findall/3}.
\begin{smallverbatim}
piped_findall(Pattern, Goals, Results) :-
	term_variables(Goals, Vars),
	pipe_create(Goals, Pattern+Vars, ID),
	pipe_all_results(ID, Results),
	stop(ID).
\end{smallverbatim}
The predicate \texttt{pipe\_all\_results/2} collects all results from the pipeline, returning them in a list.
\begin{smallverbatim}
pipe_all_results(End, Results) :-
	receive(End, In),
	(
	  In == the(done) -> Results = []
	; In == no -> pipe_all_results(End, Results)
	; the(Result+_) = In,
	  Results = [Result | Rest],
	  pipe_all_results(End, Rest)
	).
\end{smallverbatim}
Note that because goals are executed and backtracked in the order in which they occur in the conjunction, the results of \texttt{piped\_findall/3} are in the same order as those of \texttt{findall/3}.

\subsection{Identifying Pipeline Parallelisable Problems}
Conjunctions can be sped up using pipeline parallelism because in the pipeline backtracking occurs on a goal-per-goal basis.
This means that while one goal is being backtracked, other goals can simultaneously execute and generate new (partial) solutions.
Every conjunction can be executed using pipeline parallelism.
However, this does not mean that every conjunction benefits from it.

Each stage in a pipeline can be seen as a consumer, accepting partial solutions from a previous stage.
On the other hand, stages can also be seen as producers, using the partial solutions they receive to generate new (partial) solutions.
When a stage's goal is non-deterministic, we call the stage a non-deterministic producer.
Non-deterministic producers can produce \emph{more than one} new partial solution for each partial solution that they receive.

Only pipelines containing one or more non-deterministic producers have the potential to speed up execution.
In these types of pipelines, a non-deterministic producer can produce multiple partial solutions, while subsequent stages are simultaneously consuming them.
If no non-deterministic producer is present in a pipeline, each stage will only start executing when the previous stage has terminated.
Because of this, a conjunction consisting of fully (semi)deterministic subgoals, cannot be sped up by pipeline parallelism.
Note that even though a conjunction of subgoals may be (semi)deterministic, its subgoals may not be, and therefore it is still possible that pipeline parallelism results in a speedup.
Because concurrency in a pipeline is only possible \emph{starting from} the first non-deterministic producer, it is useless to create a pipeline in which the first goals are (semi)deterministic.
It is better to exclude these goals from the pipeline.

Besides taking into account the determinism of subgoals, one also needs to take into account the size of the workload represented by a subgoal.
Very small goals such as \texttt{X is Y*3} are usually very fast to execute and as such do not represent a big workload.
Therefore, it can be better to decrease the \emph{granularity} of the parallelisation by grouping these types of subgoals with other subgoals.
A method for compile-time granularity estimation can be found in \cite{Debray:1990:TGA:93542.93564}.

\section{Performance}
\label{sec:performance}
We present results from three different benchmarks, each testing one of the three types of parallelism discussed earlier: competitive or-parallelism, independent and-parallelism and pipeline parallelism.

All experiments were run on a machine with two quad-core Intel Xeon E5620 CPUs running at 2.40GHz, supporting a total of sixteen threads using HyperThreading.
The machine has a total of 24GB of memory and runs Linux 2.6.32.
Benchmarks were run with a 64-bit version of hProlog and results were compared with Logtalk version 2.42.4 with a 64-bit SWI-Prolog 5.11.18 backend.
\begin{figure}[t!]
\centering
\includegraphics{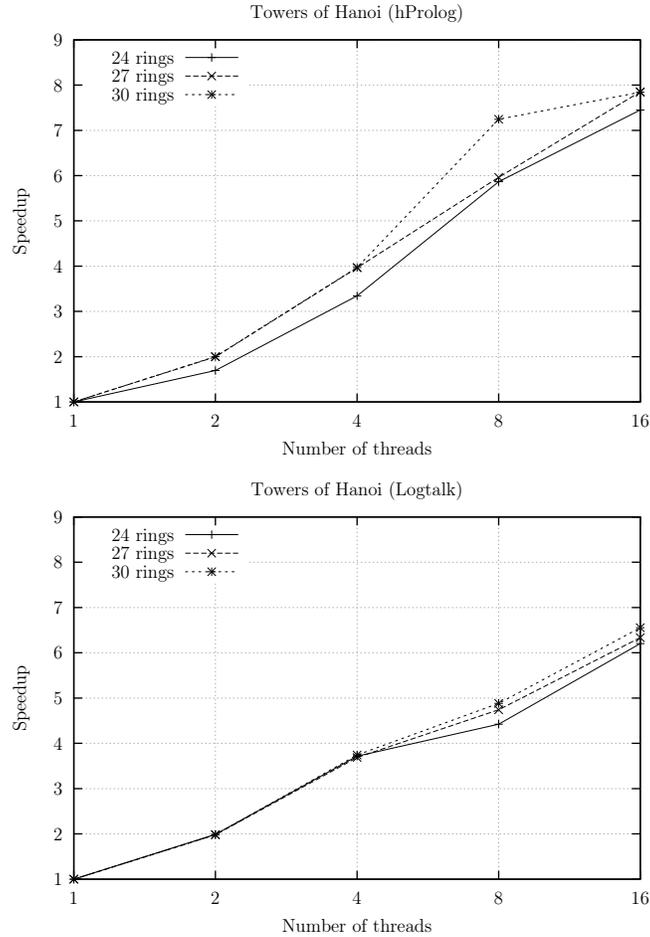}
\caption{Towers of Hanoi: speedup relative to single-threaded solution in hProlog and in Logtalk, average over 10 iterations}
\label{fig:hanoi_plot}
\end{figure}

\subsection{Independent And-Parallelism}
The \emph{Towers of Hanoi} problem is easily parallelisable using independent and-parallelism.
The benchmark solved the problem recursively and we compared the running times when using 1, 2, 4, 8 and 16 threads.

The plot in \autoref{fig:hanoi_plot} shows that our implementation scales relatively well to a high number of threads.
When using 2 and 4 threads we achieve a nearly ideal speedup, halving running times when doubling the amount of threads.
This occurs especially with a larger amount of rings.
However, the efficieny starts to decrease at 8 threads and with 30 rings the decrease at 16 threads is very drastic.
The behaviour at 16 threads is caused by the fact that there are only 8 physical processor cores but 16 hardware threads (provided by HyperThreading).
Thus, using 16 threads pushes the limits of what these hardware threads can provide.

As the plot shows, hProlog seems to be more efficient than Logtalk with SWI-Prolog as its backend, especially at higher threads counts.
Note that as a baseline, for this benchmark hProlog is on average 2 times faster than Logtalk with SWI-Prolog as its backend.
This also explains the speedups at 24 rings in hProlog: because hProlog is so fast, the relative overhead of multiple threads is bigger in hProlog than it is in Logtalk, resulting in a smaller efficiency.

\subsection{Competitive Or-Parallelism}
We used the previously mentioned water jugs problem to benchmark the performance of competitive or-parallelism.
\ifllncs
We let a hill-climbing, depth-first and breadth-first search algorithm compete to provide the solution to the problem.
\else
We let a hill-climbing (HC), depth-first (DF) and breadth-first (BF) search algorithm compete (COP) to provide the solution to the problem.
\fi
To compare the result of competitive or-parallelism with a single algorithm, the algorithms were also run separately.
The size of the jugs was 5 and 9 liters.
\begin{table}
\centering
\ifllncs
	\caption{Water jugs: slowdown compared to fastest algorithm, average of 25 iterations, with absolute running time of the fastest algorithm in seconds}
\fi
\label{tab:jugs}
\ifllncs
\begin{tabular*}{0.9\textwidth}{@{\extracolsep{\fill}} c|r r r | r r}
\else
\begin{tabular}{c|r r r | r r}
\fi
\toprule
\ifllncs
Liters & Hill Climbing & Depth First & Breadth First & Competitive & Time \\
\else
\small{Liters} & \small{HC} & \small{DF} & \small{BF} & \small{COP} & \small{Time} \\
\fi
\midrule
1 & 1.38 & 1571.11  & 1.00     & 1.78 & 0.00252 \\
2 & 1.00 & 1.86     & 1.19     & 1.02 & 2.15184 \\
3 & 1.00 & 14127.60 & 683.20   & 1.80 & 0.00020 \\
4 & 1.00 & 34055.00 & 2.50     & 3.00 & 0.00008 \\
%5 & Inf & NaN & Inf & Inf &            0.00000 \\
6 & 1.62 & 20.37    & 1.00     & 1.60 & 0.02208 \\
7 & 1.00 & 2270.60  & 58448.00 & 2.40 & 0.00020 \\
8 & 1.00 & 2036.75  & 70.50    & 1.75 & 0.00016 \\
9 & 1.00 & 7805.00  & 2.00     & 4.00 & 0.00004 \\
%10 & Inf & NaN & Inf & Inf &           0.00000 \\
11 & 14.82 & 1.00   & 3.97     & 1.17 & 0.05180 \\
12 & 1.00 & 3895.36 & 1752.50  & 1.29 & 0.00056 \\
13 & 2.41 & 89.14   & 1.00     & 1.16 & 0.00148 \\
14 & 2.14 & 133.29  & 1.00     & 2.14 & 0.00028 \\
\bottomrule
\ifllncs
	\end{tabular*}
\else
	\end{tabular}
\fi
\ifllncs
\else
	\caption{Water jugs: slowdown compared to fastest algorithm, average of 25 iterations, with absolute running time of the fastest algorithm in seconds}
\fi
\end{table}

As the results in \autoref{tab:jugs} show, the competitive or-parallel running time is always somewhat slower than the fastest algorithm run separately.
However, this slowdown is to be expected and can be ascribed to thread scheduling, the cost of creating and shutting down threads and the cost of allocating memory resources for Prolog engines for each thread.  
Most of the time, the competitive or-parallel approach is less than 2 times slower than the fastest algorithm.
However, in general, the competitive or-parallel approach is considerably faster than the slowest algorithm.

In the case of 9 liters, the competitive or-parallel approach is 4 times slower than the hill climbing algorithm, while the breadth first algorithm is just 2 times slower.
A similar event occurs at 4 and 1 liters.
These occurrences are probably explained by the overhead of multi-threading and the small running times of the fastest algorithm.

The results for this benchmark are very similar to those obtained by running the same benchmark using Logtalk.
Logtalk also exhibits similar behaviour at 4 and 9 liters.

\subsection{Pipeline Parallelism}
Finding the intersection of a number of sets is a problem that is not natural to parallelise using independent and- or competitive or-parallelism.
However, using pipeline parallelism, one can achieve considerable speedups, depending on the number of sets and their size.
We compared the performance of the regular \texttt{findall/3} predicate to that of the previously described \texttt{piped\_findall/3} predicate.

To calculate the intersection between \emph{n} sets, we used conjunctions of the form
\begin{smallverbatim}
member(X, L1), member(X, L2), \ldots, member(X, Ln)
\end{smallverbatim}
where L1, L2, ..., L\emph{n} are lists of a given length.
For example, we used
\ifllncs
\begin{smallverbatim}
?- findall(X, (member(X, L1), member(X, L2)), R).
?- piped_findall(X, (member(X, L1), member(X, L2)), R).
\end{smallverbatim}
\else
\begin{smallverbatim}
?- findall(X,
     (member(X, L1), member(X, L2)), R).
?- piped_findall(X,
     (member(X, L1), member(X, L2)), R).
\end{smallverbatim}
\fi
to compare the performance of finding the intersection of two sets.
We tested the performance of both predicates consecutively using 2, 3, 4, \ldots, 16 sets.
Note that to exploit the pipeline to its fullest extent, we have opted not to use \texttt{memberchk/2} for the goals following the first one, using \texttt{member/2} instead.

When all sets are equal, checking every element results in a full traversal of the pipeline, maximising its use.
We used this scenario as a best-case benchmark.
When the sets do not share any element, the second \texttt{member/2} in the conjunction fails immediately, minimising the use of the pipeline.
We used this scenario as a worst-case benchmark.

\ifllncs
\begin{figure}[b!]
\else
\begin{figure}[t!]
\fi
\centering
\includegraphics{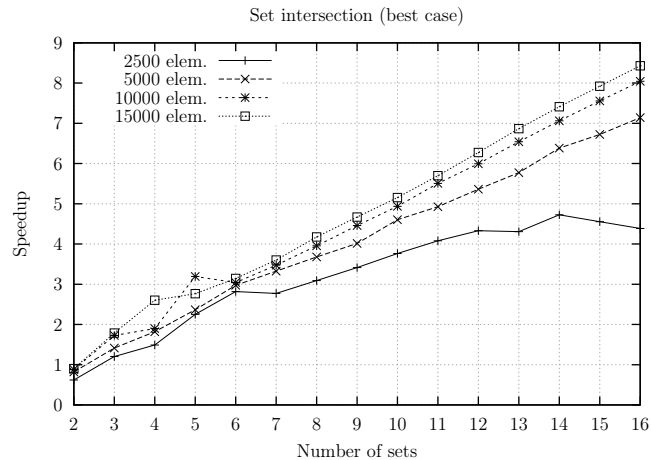}
\caption{Set intersection best case: average speedup compared to regular findall over 25 iterations}
\label{fig:intersection_best}
\end{figure}
The plot for the best-case scenario in \autoref{fig:intersection_best} shows that considerable speedups can be achieved by using \texttt{piped\_findall/3}.
Starting at three or more sets, \texttt{piped\_findall/3} is consistently faster than \texttt{findall/3}.
Overall, speedups increase as the size of the sets or the number of sets increases.
However, note that in the case of sets of 2500 elements, the size of the speedup peaks at around 14 threads.
Past this point, the overhead incurred by pipelining probably starts to exceed the speedup gained from using more threads.

\ifllncs
\begin{figure}[t!]
\else
\begin{figure}[t!]
\fi
\centering
\includegraphics{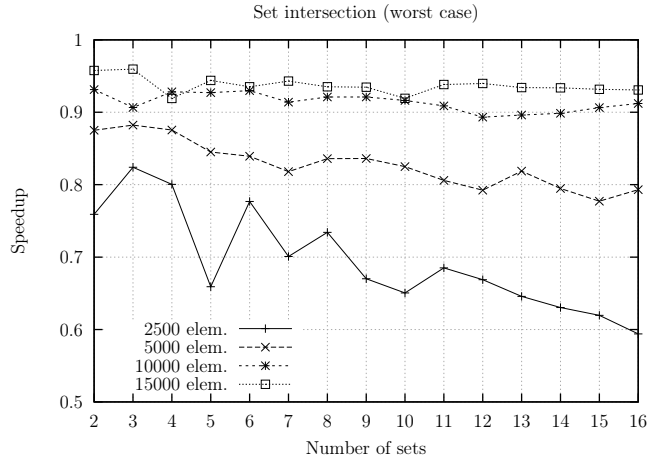}
\caption{Set intersection worst case: average speedup compared to regular findall over 25 iterations}
\label{fig:intersection_worst}
\end{figure}
The plot for the worst-case scenario in \autoref{fig:intersection_worst} shows that there is a significant slowdown when the main benefit of the pipeline is not used.
The smaller the sets, the bigger the slowdown becomes.
Increasing the number of sets also increases the slowdown.
This is because increasing the number of sets extends the length of the pipeline and thus more threads need to be created.
However, since partial solutions never pass the second thread, all other threads are unused and only create more overhead.

\section{Related Work}
\label{sec:related_work}
There are a number of Prolog compilers implementing multi-threading support on different levels.
Among them are Logtalk, SWI-Prolog and LeanProlog, all of which implement support for multi-threading in different ways.
They all support independent and-parallelism and competitive or-parallelism either directly or indirectly.
However, as far as we know, none of them include direct support for parallelism through pipelining, although we expect this can be implemented using existing primitives.
\subsection{Logtalk}
Logtalk is ``an object-oriented programming language that can use most Prolog implementations as a back-end compiler''.
As such, it is not a Prolog compiler in the strict sense, but rather a layer on top of an existing compiler.
Nevertheless, Logtalk provides an interesting comparison since it provides a number of high-level predicates to leverage multi-threading.
One of these predicates was already mentioned earlier: \texttt{threaded/1}.
\texttt{threaded/1} can be used for both independent and-parallelism and competitive or-parallelism.

\begin{sloppypar}
Logtalk also provides a few high-level predicates to concurrently execute single goals.
These are \texttt{threaded\_call(Goal)} and \texttt{threaded\_once(Goal)}, which call the given goal in a separate thread.
The results from these goals can be accessed using \texttt{threaded\_exit(Goal)} and \texttt{threaded\_peek(Goal)}.
\texttt{threaded\_exit/1} waits for a result to become available and unifies \texttt{Goal} with that result.
\texttt{threaded\_peek/1} works the same way, but fails if no result is available, instead of waiting.

Note that \texttt{threaded\_call/1} will generate a single solution to its goal in a new thread and then suspend itself.
Only after \texttt{threaded\_exit/1} has consumed a solution will the next solution be generated.
This is different from our approach, where a thread is not suspended after it has generated a solution.

Logtalk also supports one-way asynchronous calls with \texttt{threaded\_ignore(Goal)}, which executes the goal in a thread and then succeeds.
\end{sloppypar}

Since Logtalk can use a number of Prolog compilers as its backend, all multi-threading predicates in Logtalk are implemented on top of a low-level API, specified by the ISO standardisation proposal for multi-threading support in Prolog \cite{ISO/IEC-DTR-13211-5:2007}.

\subsection{ISO Standardisation Proposal for Multi-Threading Support in Prolog}
This ISO standardisation proposal is based on the design found in SWI-Prolog \cite{DBLP:conf/iclp/Wielemaker03}.
The predicates have also been implemented in a number of other Prolog systems, including XSB \cite{xsb_prolog_manual} and YAP \cite{yap_manual}.

The low-level multi-threading predicates described in the proposal are based on the semantics of POSIX threads.
The predicates form a comprehensive set, supporting the creation and destruction of threads, message queues, and mutexes.
The predicates for creating threads accept various options.
Some of these options allow specifying low-level details of thread creation, such as the limit to which the global stack, local stack, C stack or trail can grow.
For more details, we point the reader to \cite{ISO/IEC-DTR-13211-5:2007}.

Compared to the predicates in this API, our language constructs are more high-level.
They do not allow such fine-grained control over limits of memory areas.
Also, we do not support mutexes (yet).
Our goal was to create a high-level interface to multi-threading which still allows a relatively high degree of control.
As such, we intentionally separated the semantics of POSIX threads from our interface.

\subsection{SWI-Prolog}
On top of the low-level multi-threading support, SWI-Prolog also provides two higher-level predicates for using independent and-parallelism and competitive or-parallelism \cite{Wielemaker:swipl/library/thread.pl}:
\begin{smallverbatim}
concurrent(N, Goals, Options)
first_solution(X, Goals, Options)
\end{smallverbatim}
\texttt{concurrent/3} accepts a list of independent goals and executes them concurrently using \texttt{N} threads.
Contrast this to Logtalk's implementation of \texttt{threaded/1}, where the number of threads to use cannot be specified.
\texttt{first\_solution/3} calls a list of goals and uses the first result that is calculated.
\texttt{X} functions as the answer pattern, specifying what variables need to be returned.

The variable \texttt{Options} is a list of options that is passed to the low-level predicates that create threads.

\subsection{LeanProlog}
Out of all Prolog compilers that we have mentioned, LeanProlog's \cite{Tarau:2011:CPC:1926354.1926364} high-level multi-threading support most closely matches ours.
Various aspects of our interface originated from ideas implemented in LeanProlog and its predecessor BinProlog \cite{DBLP:journals/corr/abs-1102-1178}.
We have also implemented the \texttt{multi\_fold/3} predicate described in \cite{Tarau:2011:CPC:1926354.1926364} using our language constructs.
However, for the sake of brevity we have not included this implementation here.

LeanProlog's support is focused on an Interactor API.
LeanProlog supports what we call threads linked to a hub.
By default, threads share the code zone but have separate symbol tables.
LeanProlog supports symbol garbage collection and separate symbol tables allow garbage collection to occur safely in multiple threads, without the need for synchronisation.
One can optionally specify whether the code zone must also be cloned.

\section{Conclusion and Future Work}
\label{sec:conclusion}
We have shown that our interface provides the means to build high-level multi-threading constructs and that our implementation's performance is comparable to that of Logtalk 2.42.4 on top of SWI-Prolog 5.11.18.
We have discussed two common types of high-level parallelism and have identified some of their limitations.
Based on these observations, we have proposed a new type of parallelism, pipeline parallelism, and we have identified the types of problems where this type of parallelism is suitable.

As for future work, we plan to search for real-world problems for which pipeline parallelism is useful.
\emph{Implicit} pipeline parallelism with automatic granularity analysis is another topic we would like to research.
Furthermore we would like to address some of the limitations of our current design as outlined in \autoref{subsec:limitations}.
We would also like to implement better support for exceptions and make better use of them by using them in our language constructs to signal errors.

\section*{Acknowledgements.}
\ifllncs
\else
Thanks to Bart Demoen for guiding me during my research and providing me with the source code to hProlog.
\fi
Thanks to Paulo Moura for his interesting conversations, his useful insights and Logtalk's set of multi-threading benchmarks.
Thanks to Paul Tarau for his earlier work on multi-threaded logic engines and his helpful comments concerning our interface.

\bibliographystyle{splncs03}
\bibliography{main}{}

\end{document}